\documentclass{article}
\usepackage{spconf,amsmath,graphicx}
\usepackage{amsfonts}%
\usepackage{graphicx}
\usepackage{subcaption}
\usepackage{booktabs}
\usepackage{multirow}
\usepackage{placeins}
\usepackage{url}
\usepackage{tikz}
\usepackage{amssymb}
\usepackage{pifont}

\usepackage{xcolor}
\usepackage{comment}
\usepackage[super]{nth}

\usepackage{etoolbox}
\newcommand{\zerodisplayskips}{%
  \setlength{\abovedisplayskip}{0pt}%
  \setlength{\belowdisplayskip}{0pt}%
  \setlength{\abovedisplayshortskip}{0pt}%
  \setlength{\belowdisplayshortskip}{0pt}}
\appto{\normalsize}{\zerodisplayskips}
\appto{\small}{\zerodisplayskips}
\appto{\footnotesize}{\zerodisplayskips}

{\fontsize{9}{9}}

\def\xvec{\mathbf{x}}

\def\Xmat{\mathbf{X}}

\def\Amat{\mathbf{A}}
\def\avec{{\mathbf a}}

\def\gen{\mathrm{gen}}
\def\rec{\mathrm{rec}}

\def\S{\mathrm{S}}
\def\T{\mathrm{T}}

\title{Low-Resource Domain Adaptation for Speaker Recognition Using Cycle-GANs}
\name{Phani Sankar Nidadavolu, Saurabh Kataria, Jes\'us Villalba, Najim Dehak}
\address{Center for Language and Speech Processing, Johns Hopkins University, Baltimore, MD, USA\\
{\small\tt\{snidada1,skatari1,jvillal7,ndehak3\}@jhu.edu}}
\begin{document}
%
\maketitle
\begin{abstract}
Current speaker recognition technology provides great performance with the x-vector approach. However, performance decreases 
when the evaluation domain is different from the training domain, an issue usually addressed with domain adaptation approaches.
Recently, unsupervised domain adaptation using cycle-consistent Generative Adversarial Networks (CycleGAN) has received a lot of attention. 
CycleGAN learn mappings between features of two domains given non-parallel data. 
We investigate their effectiveness in low resource scenario i.e. when limited amount of \emph{target} domain data is available for adaptation, a case unexplored in previous works. We experiment with two adaptation tasks: microphone to telephone and a novel reverberant to clean adaptation with the end goal of improving speaker recognition performance. Number of speakers present in \emph{source} and \emph{target} domains are 7000 and 191 respectively. By adding noise to the \emph{target} domain during CycleGAN training, we were able to achieve better performance compared to the adaptation system whose CycleGAN was trained on a larger \emph{target} data. 
On reverberant to clean adaptation task, our models improved EER by 18.3\% relative on VOiCES dataset compared to a system trained on clean data. They also slightly improved over the state-of-the-art Weighted Prediction Error (WPE) de-reverberation algorithm.

\end{abstract}

\begin{keywords}
Domain Adaptation, CycleGAN, Low Resource, Microphone-Telephone, Speaker Recognition
\end{keywords}

\section{Introduction}
\label{sec:intro}
Speaker recognition technology has made great progress in the last decade. The x-vector approach~\cite{snyder2018x} is the current state-of-the-art in this field,
providing superior performance in NIST SRE, Speakers In The Wild (SITW)~\cite{mclaren2016speakers} and VoxCeleb datasets~\cite{nagrani2017voxceleb}. x-vectors is a data-hungry approach, i.e., it requires a huge amount of labeled data ($\sim10$k speakers with multiple recordings per speaker) to be properly trained. Most data available for training consists of English speech of moderately good quality. When we apply x-vector networks trained on these data to other acoustic domains,
performance dramatically drops. We can witness this in the latest NIST SRE18~\cite{villalbajhu}. Acquiring labeled data from thousands of speakers of each domain would be very expensive or just unfeasible. However, in some cases, it is possible to get access to certain amount of data from the domain of interest denoted from now on as the \emph{target} domain. Then, we can make use of domain adaptation techniques to adapt features or models trained on data from the \emph{source} domain--the domain with plenty of training data-- to the \emph{target} domain or vice-versa. The final goal is to improve the performance on evaluation corpora sampled from the \emph{target} domain.

In this paper, we are interested in the particular case of low-resource unsupervised domain adaptation. It is low-resource because we assume a limited amount of \emph{target} domain (LT) data is available to train the adaptation system. Furthermore, it is unsupervised in two ways. First, we assume that the adaptation data doesn't have any speaker labels. Second, we assume that we don't have any paired data between \emph{source} and \emph{target} domains. By paired data, we mean, for example, the case where somebody records the same conversation through close-talk and far-field microphones at the same time. Then, we can use the matched pairs to learn mapping functions between domains.
This setting is of interest to us because this opens up the possibility of maximally taking advantage of small development set found in real data, and not use simulated sets (as done commonly in practice).

Typically, we find two types of strategies for domain adaptation. The first one consist of adapting a model trained on the \emph{source} domain to the \emph{target} domain. The second one, which is the one we follow, consist of mapping the acoustic input features from the \emph{target} domain back to the \emph{source} domain. Thus, we can use the original \emph{source} domain speaker recognition system without any changes. 

For this unsupervised scenario, CycleGAN~\cite{zhu2017unpaired} is an effective solution. Training objective function for CycleGAN is a combination of reconstruction and adversarial losses~\cite{goodfellow2014generative}. CycleGAN was first proposed in computer vision literature for image-to-image translation with non-parallel data. 
In speech research, CycleGAN has been used for mapping noisy speech to clean speech, improving automatic speech recognition (ASR) trained on clean speech~\cite{mimura2017cross,meng2018cycle}, voice conversion~\cite{kaneko2017parallel,fang2018high,kameoka2018stargan}, gender adaptation~\cite{hosseini2018multi}, and microphone to telephone (\textit{mic-tel}) adaptation for speaker recognition~\cite{nidadavolu2019cycle}.

All works cited above have demonstrated the capability of CycleGAN for unsupervised domain adaptation across wide range of tasks. However, they were trained on comparable amounts of \emph{source} and \emph{target} domain data. For example, in the noisy to clean adaptation work in~\cite{meng2018cycle}, the same amount of data is used for both domains- \emph{target} domain data is created by artificially adding noise to clean data. Also in~\cite{kaneko2017parallel}, the authors create \emph{target} domain data by augmenting real noisy data with data created by artificially adding noise to the \emph{source} domain. In our previous work ~\cite{nidadavolu2019cycle}, we improved the performance of the speaker recognition system trained on telephone corpus on Speakers In The Wild (SITW)~\cite{mclaren2016speakers}, a microphone corpus. For that, we used development portion of SITW and the much larger VoxCeleb1 dataset~\cite{nagrani2017voxceleb} as the \emph{target} domain data to learn the feature mapping function.


In this work, we experimented with \textit{mic-tel} (as done in~\cite{nidadavolu2019cycle}) and a novel reverberant to clean speech (\textit{reverb-clean}) adaptation tasks, both with the end goal of improving speaker recognition performance under the low-resource setting. For both tasks, CycleGANs were trained with data from ~7000 speakers from the \emph{source} domain and 191 speakers from the \emph{target} domain. We observed that the low-resource setting limits the capability of CycleGAN to learn meaningful mappings between domains. The drop in performance can be attributed to the phenomenon of over-fitting or lack of sufficient variation between the domains. 
To overcome this disadvantage, we experimented with artificially adding noise to the LT data before training CycleGAN. The motivation behind adding noise is to prevent over-fitting by making it act as a regularizer and to make both the \emph{source} and \emph{target} distributions more different (telephone corpus is often considered as clean) of which the adversarial loss can take advantage.
By adding noise, adaptation system with CycleGAN trained on LT data performed slightly better than adaptation system whose CycleGAN was trained on larger amounts of \emph{target} domain data. We observed this on both the tasks. More importantly, we observed that noise addition speeds up training (slightly better performance with fewer epochs of training). For the \textit{reverb-clean} task, CycleGAN trained with LT data and added noise yielded good improvements (18.3\% relative improvement in EER) w.r.t the baseline speaker recognition system trained on clean and tested on reverberant speech. CycleGAN adaptation also showed slight improvements over the state-of-the-art weighted prediction error (WPE)~\cite{nakatani2010speech} de-reverberation algorithm.

\section{Low-resource C\MakeLowercase{ycle}GAN System}
\label{sec:cg_system}

The adaptation procedure in this study is as follows. We start-off by training an x-vector network~\cite{snyder2018x} on \emph{source} domain data. 
During the evaluation, we map the input features of the x-vector network from \emph{target} domain data to the \emph{source} domain. 
This is accomplished by learning feature mapping functions between the two domains.
The training corpora to learn the feature mapping functions consists of audio samples $\Amat_{\S}=\{\avec_{\S,i}\}_{i=1}^{N}$ and $\Amat_{\T}=\{\avec_{\T,i}\}_{i=1}^{M}$ drawn from two different domains: \emph{source} $S$ and \emph{target} $T$ with distributions $\avec_{\S,i}\sim q_{\S}(\avec)$ and $\avec_{\T,i}\sim q_{\T}(\avec)$ respectively. In this work we assume $M << N$. Noise is then added to the \emph{target} domain audio samples which result in a transformed \emph{target} domain distribution $T'$. Log mel-filter bank features are extracted from audio samples of \emph{source} and transformed \emph{target} domain distributions denoted as $\Xmat_{\S}=\{\xvec_{\S,i}\}_{i=1}^{N}$ and $\Xmat_{\T'}=\{\xvec_{\T',i}\}_{i=1}^{M}$. The distributions in filter bank space are $\xvec_{\S,i}\sim p_{\S}(\xvec)$ and $\xvec_{\T',i}\sim p_{\T'}(\xvec)$ respectively.
$\Xmat_{\S}$ and $\Xmat_{\T'}$ are used as features from two different distributions to train the feature mapping functions.
Speaker labels from either domain are not needed to train the feature mapping function. The training data is assumed non-parallel which makes the adaptation procedure unsupervised. Noise is added to the \emph{target} domain audio only during training. During evaluation original audio samples of the evaluation data sampled from the \emph{target} domain data are used to extract filter bank features which are then mapped to the \emph{source} domain. The \emph{target} domain data used to train and evaluate the feature mapping system has no speaker overlap.

\begin{figure}[t]
\begin{minipage}[b]{1.0\linewidth}
  \centering
  \centerline{\includegraphics[width=8.0cm]{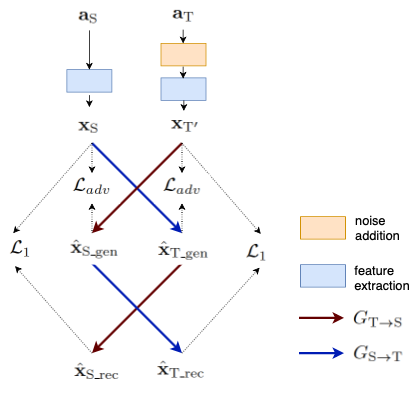}}
\end{minipage}
\caption{CycleGAN layout}
\label{fig:cycle_gan_layout}
\vspace{-4mm}
\end{figure}

The feature mapping between domains is achieved by training a CycleGAN system. The training procedure for CycleGAN with LT data is outlined in Figure \ref{fig:cycle_gan_layout}.
It consists of two generators, each of which maps features of one domain to the opposite. It also has two discriminators, one for each domain (not shown in Figure~\ref{fig:cycle_gan_layout}). Each discriminator tries to distinguish between  real features from its domain and features mapped (generated) from the opposite 
domain to that domain.

\subsection{CycleGAN training objectives}
\label{cg_training}
The training procedure is as follows: two batches of features $\xvec_{\S}$ and $\xvec_{\T'}$ are sampled from \emph{source} and transformed \emph{target} domains respectively. The generator $G_{\T \rightarrow \S}$ maps $\xvec_{\T'}$ to \emph{source} domain, producing features $\hat{\xvec}_{\S\_\gen}$. A discriminator $D_{\S}$ is trained to discriminate between original and generated \emph{source} domain features. The generator $G_{\T \rightarrow \S}$ is then trained to output features $\hat{\xvec}_{\S\_\gen}$ that appear identical to the original \emph{target} domain features $\xvec_{\S}$. 
This is accomplished by training the generator with an adversarial loss~\cite{goodfellow2014generative}. As recommended in \cite{mao2017least}, we used the least-squares objective to train our generator and discriminator (LS-GAN). $D_{\S}$ and $G_{\T \rightarrow \S}$ are trained to minimize the objectives \ref{disc_obj} and \ref{adv_obj} respectively. 

\begin{align} 
        \label{disc_obj}
            L_{\mathrm{disc}}(G_{\T \rightarrow \S}, D_{\S}, \Xmat_{\T'},\Xmat_{\S}) = &\mathop{\mathbb{E}_{\xvec \sim p_{\T'}}[D_{\S}(G_{\T \rightarrow \S}(\xvec))^2]}
            +\nonumber \\
            &\mathop{\mathbb{E}_{\xvec \sim p_{\S}}[(D_{\S}(\xvec)-1)^2]} 
\end{align}
\begin{align} 
\label{adv_obj}
            L_{\mathrm{adv}}(G_{\T \rightarrow \S}, D_{\S}, \Xmat_{\T'}) = \mathop{\mathbb{E}_{\xvec \sim p_{\T'}}[(D_{\S}(G_{\T \rightarrow \S}(\xvec))-1)^2]}
\end{align}

Equivalently, the other generator-discriminator ($G_{\S \rightarrow \T}$, $D_{\T}$) pair is trained in a similar fashion to transfer features from \emph{source} domain to transformed \emph{target} domain. During evaluation, since we map features from the \emph{target} domain to the \emph{source} domain without adding noise, we keep the notation of generators as $G_{\T \rightarrow \S}$ instead of $G_{\T' \rightarrow \S}$. 

A single generator-discriminator pair trained with adversarial loss would suffice, in theory, to transfer features from one domain to the opposite domain. 
However, this leads to an ill-posed problem with adversarial loss putting a weak constraint on the generators. Thus, the generator could create many possible features which appear to be drawn from the true distribution but may fail to preserve the structure present in the signal like the linguistic information, speaker and gender information. To restrict the space of possible mappings from the generator, CycleGAN enforce cycle-consistency constraint on the generators - reconstructing the original features, e.g. $\xvec_{\S}$, from the generated features
in the opposite domain, e.g., $\hat{\xvec}_{\T\_\gen}$. This is achieved by minimizing the objective in~\eqref{eq3} between $\xvec_{\S}$ and $\hat{\xvec}_{\S\_\rec}=G_{\T \rightarrow \S}(\hat{\xvec}_{\T\_\gen})$ where we used $l_1$ distance as the metric. We refer to this loss as forward cycle consistency loss. Similarly, the loss computed between $\xvec_{\T'}$ and $\hat{\xvec}_{\T\_\rec}$ is referred to as backward cycle consistency loss. The final cycle consistency loss is the combination of both these objectives and is given in~\eqref{eq4}.

\begin{align} 
            \label{eq3}
            &L_{1}(G_{\S \rightarrow \T},G_{\T \rightarrow \S}, \Xmat_{\S}) = \mathop{\mathbb{E}_{\xvec \sim p_{\S}}{||G_{\T \rightarrow \S}(G_{\S \rightarrow \T}(\xvec))-\xvec||}_1}   
\end{align}

\begin{align} 
\label{eq4}
            L_{\mathrm{cyc}}(G_{\S \rightarrow \T},G_{\T \rightarrow \S}, \Xmat_{\S}, \Xmat_{\T'}) &= L_{1}(G_{\S \rightarrow \T},G_{\T \rightarrow \S}, \Xmat_{\S}) \nonumber\\
            &+ L_{1}(G_{\T \rightarrow \S},G_{\S \rightarrow \T}, \Xmat_{\T'})
\end{align}
        
Finally, both the generators of CycleGAN are trained using multi-task objective: by minimizing both the adversarial and cycle consistency objectives as shown in~\eqref{eq5}. 
${\lambda}_{\mathrm{cyc}}$ and ${\lambda}_{\mathrm{adv}}$ in~\eqref{eq4} denote the weights assigned to cycle consistency loss and adversarial loss respectively. 

\begin{align} 
\label{eq5}
            &L(G_{\T \rightarrow \S}, G_{\S \rightarrow \T}, D_{\T}, D_{\S}, \Xmat_{\S}, \Xmat_{\T'}) \nonumber \\
            &= {\lambda}_{adv} L_{\mathrm{adv}}(G_{\T\rightarrow \S}, D_{\S}, \Xmat_{\T'}) \nonumber\\
            \quad & + {\lambda}_{\mathrm{adv}} L_{\mathrm{adv}}(G_{\S\rightarrow \T}, D_{\T}, \Xmat_{\S}) \nonumber\\
            \quad & + {\lambda}_{\mathrm{cyc}} L_{\mathrm{cyc}}(G_{\S \rightarrow \T}, G_{\T \rightarrow \S}, \Xmat_{\S}, \Xmat_{\T'})
\end{align}

\begin{table}
	\caption{Generator architecture}
    \label{tab:arch_gen}

	\centering
	\vspace{-3mm}
    \resizebox{7.5cm}{!}{
	\begin{tabular}{@{}lcl@{}}
    \toprule
    \textbf{Layer} & \textbf{Kernel size} & \textbf{Output} \\
    \midrule 
	Shortcut & - & h x w x 1 \\
    \midrule 
    \textbf{Downsampler} \\
    Conv, ReLU  & [3,3,1] & h x w x 32 \\
    Conv, IN, ReLU  & [3,3,2] & h/2 x w/2 x 64 \\
    Conv, IN, ReLU  & [3,3,2] & h/4 x w/4 x 128 \\
    (ResBlock, ReLU) x 9& - & h/4 x w/4 x 128 \\
    \midrule 
    \textbf{Upsampler} \\
    Deconv, IN, ReLU & [3,3,2] & h/2 x w/2 x 64 \\
    Deconv, IN, ReLU & [3,3,2] & h x w x 32 \\
    Conv & [3,3,1] & h x w x 1 \\
	\midrule 
	Addition & - & h x w x 1 \\
	\bottomrule
    \end{tabular}}
    
\end{table}

\begin{table}
	\caption{ResBlock architecture}
    \label{tab:arch_resnet}
\vspace{-3mm}
	\centering
    \resizebox{7.5cm}{!}{
	\begin{tabular}{@{}lcl@{}}
    \toprule
    \textbf{Layer} & \textbf{Kernel size} & \textbf{Output} \\
    \midrule 
	Shortcut & - & h/4 x w/4 x 128 \\
    \midrule 
    Conv, IN, ReLU  & [3,3,1] & h/4 x w/4 x 128 \\
    Conv, IN  & [3,3,1] & h/4 x w/4 x 128 \\
    \midrule 
    Addition  & - & h/4 x w/4 x 128 \\
	\bottomrule
    \end{tabular}}
    \vspace{-6mm}
\end{table}

\begin{table}
	\caption{Discriminator architecture}
    \label{tab:arch_discnet}
\vspace{-3mm}
	\centering
    \resizebox{7.5cm}{!}{
	\begin{tabular}{@{}lcl@{}}
    \toprule
    \textbf{Layer} & \textbf{Kernel size} & \textbf{Output} \\
    \midrule 
    Conv, LeReLU  & [4,4,2] & h/2 x w/2 x 64 \\
    Conv, LeReLU  & [4,4,2] & h/4 x w/4 x 128 \\
    Conv, LeReLU  & [4,4,2] & h/8 x w/8 x 256 \\
    Conv, LeReLU  & [4,4,1] & h/8 x w/8 x 512 \\
    Conv & [4,4,1] &  h/8 x w/8 x 1 \\
	\bottomrule
    \end{tabular}}
    \vspace{-3mm}
\end{table}

\vspace{-2mm}
\subsection{Network architectures}
\label{cg_archs}
The architecture for generators is given in Table \ref{tab:arch_gen}. It is a full-convolutional network with a downsampler-upsampler architecture. Kernel size is described as {[\textit{kernel\_h}, \textit{kernel\_w}, \textit{stride}]}. Input shape to the network is \textit{h} x \textit{w} and output shape of each layer is \textit{h} x \textit{w} x \textit{channels}. Similar to the work in \cite{nidadavolu2019cycle}, we used a short cut connection from input and added the shortcut to the output of the last convolutional layer of the network. The architecture for the residual block used in the generator is given in Table \ref{tab:arch_resnet} and the architecture for discriminators is given in Table \ref{tab:arch_discnet}. The slopes of all Leaky ReLU (LeReLU) functions were set to 0.2. Since we used least squares objective to train the discriminator, we have not applied any non-linear activation at the output.
For both \textit{mic-tel} and \textit{reverb-clean} adaptation tasks, we used same network architectures and same training objectives

\section{Experimental details for \emph{\MakeLowercase{mic-tel}} adaptation}
\label{sec:exp_details_tel_mic}
In this section, we describe the experimental setup for low-resource \textit{mic-tel} adaptation.  
\vspace{-3mm}

\subsection{\emph{source} and \emph{target} domain datasets}
\label{ssec:tel_mic_datasets}
Telephone domain data (\emph{source} domain) used to train x-vector system consisted of recordings from datasets SRE04-10, Mixer6 and Switchboard 1-Phase 1, 2, and 3. This gave us 90946 utterances from 6986 speakers. This was also used as the \emph{source} domain data to train the CycleGAN system. Development corpus of SITW (referred to as SITW \textit{dev}) was used as microphone corpus from the \emph{\emph{target} domain} to train the CycleGAN system. It has 4439 utterances from 119 speakers. The mismatch in the number of speakers and utterances between both the domains can clearly be observed justifying the need for low-resource domain adaptation. SITW evaluation corpus (referred to as SITW \textit{eval}) was used to evaluate the system. No speaker overlap exists between SITW \textit{eval} and \textit{dev} corpora. The microphone speech was down-sampled to 8KHz to match the sampling frequency of telephone speech. 
\vspace{-3mm}
\subsection{Noise addition to the \emph{target} domain data}
\label{ssec:tel_mic_noise_addition}
We added noise to the speech signals from the \emph{target} domain to train CycleGAN. To add noise we used 930 "noise" samples from MUSAN \cite{snyder2015musan} corpus. Noise was added as foreground noises at the interval of 1 second with the signal to noise ratios (SNRs) ranging from 0 to 15dB. The "music" and "babble" portions of MUSAN corpus were not used in this work. 
Noise addition was done only on \emph{target} domain data during the training of CycleGAN system. The original \emph{target} domain data (without noise) was not used during training. While forward passing the SITW \textit{eval} features through CycleGAN, no noise was added. No noise was added to the \emph{source} domain data during the training of x-vector system and CycleGAN system.  

\vspace{-3mm}
\subsection{Baseline system}
\label{sec:tel_mic_xvec_system}
The x-vector system was based on Kaldi~\cite{povey2011kaldi}. We used the same setup as in SRE16 Kaldi recipe\footnote{\url{https://github.com/kaldi-asr/kaldi/tree/master/egs/sre16/v2}} but without any data augmentation. The x-vector system was trained on telephone corpus mentioned in Section~\ref{ssec:tel_mic_datasets} and evaluated on SITW \textit{eval}. 
The system used 40-dimensional log Mel filter-bank features with short-time centering (300 frames). Energy-based VAD was applied to remove the non-speech frames. x-vector network was trained for 3 epochs. After the training, the network was used to extract x-vectors (speaker embeddings) for the training and evaluation corpus. The x-vectors were centered, projected to 150 dimensions using Linear Discriminant Analysis (LDA) and length normalized. Full-rank Probabilistic Linear Discriminant Analysis (PLDA)~\cite{brummer2010speaker} was used to get the scores.
Finally, scores were normalized using adaptive symmetric norm (S-Norm)~\cite{brummer2009agnitio}. In the baseline system, both the x-vector network and PLDA backend were trained on telephone speech and tested on microphone speech.

\vspace{-3.5mm}
\subsection{Adaptation system}
\label{sec:tel_mic_adapt_system}

For the adaptation system, the x-vector network and PLDA were same as in the baseline. Feature adaptation was done in the evaluation stage. First, the filter-bank features of the evaluation corpus were mapped from microphone to telephone domain by forward passing through the $G_{\T\to\S}$ generator of CycleGAN network. Then, the mapped features were used to extract the x-vectors for the evaluation data. 

\vspace{-4mm}
\subsection{CycleGAN training}
\label{ssec:tel_mic_cg_system}

Similar to x-vector system, CycleGAN system was trained on 40-dimensional log mel-filter bank features with short time centering. Energy VAD was applied on centered features to remove the non-speech frames. Two batches of features were sampled randomly from \emph{source} and transformed \emph{target} domain (\emph{target} domain with noise) during each training step. Since no parallel data exists between both the domains, the batches were drawn in a completely random fashion. The size of the batches was set to 32 and the number of contiguous frames sampled from each utterance (sequence length) was set to 127. 
The model was trained for 50 epochs. Each epoch was set to be complete when all the telephone utterances have appeared once in that epoch. Adam Optimizer was used with momentum $\beta_1=0.5$ as suggested by~\cite{radford2015unsupervised}. The learning rates for the generators and discriminators were set to 0.0003 and 0.0001 respectively. The learning rates were kept constant for the first 15 epochs and, then, linearly decreased until they reach the minimum learning rate (1e-6).
The cycle loss weight was set to 2.5 and adversarial loss weight was set to 1.0. We used PyTorch for the CycleGAN implementation.

\begin{table}
	\caption{Results for \emph{Adaptation system LT} trained without noise}
    \label{tab:tel_mic_results_baselines}
\vspace{-2mm}
	\centering
    \resizebox{\columnwidth}{!}{
	\begin{tabular}{@{}lcccc@{}}
    \toprule
    & \multicolumn{2}{c}{\textbf{SITW Core-Core}} & \multicolumn{2}{c}{\textbf{SITW Assist-Multi}} \\
    \cmidrule(lr){2-3}
    \cmidrule(lr){4-5}
    & \textbf{EER} & \textbf{DCF} & \textbf{EER} & \textbf{DCF} \\
    \midrule 
    \textbf{\emph{Baseline system S}} & 10.14 & 0.6842 & 12.72 & 0.6941 \\
    \midrule
    \textbf{\emph{Adaptation system}} &  \textbf{8.87} & \textbf{0.6548} & \textbf{10.78} & \textbf{0.6643} \\
    \textbf{\emph{Adaptation system LT}} & 9.51 & 0.6608 & 11.43 & 0.6683 \\
    \midrule
    \textbf{\emph{Baseline system S \& T}} & 7.90 & 0.6226 & 10.14 & 0.6418 \\
	\bottomrule
    \end{tabular}}
    \vspace{-3mm}
\end{table}

\section{Results for \emph{\MakeLowercase{mic-tel}} adaptation}
\label{sec:results_tel_mic_domain_adapt}
We report our results using metrics Equal Error Rate (EER) in \% and DCF (Detection Cost Function) \cite{brummer2006application} under two testing conditions of SITW corpus: \emph{Core-Core} and \emph{Assist-Multi} \cite{mclaren2016speakers}. We refer to the adaptation system trained with LT data as \emph{Adaptation system LT}.

\vspace{-3mm}

\subsection{\emph{Adaptation system LT} trained without noise}
\label{ssec:results_ltd_tgt_data_tel_mic}

First, we present results for CycleGAN trained without noise addition to the \emph{target} domain.
Table~\ref{tab:tel_mic_results_baselines} presents the results for baseline system trained on 
\emph{source} domain (\emph{Baseline system S}) and \emph{Adaptation system LT} without noise addition.
We also trained a baseline x-vector system on both \emph{source} and \emph{target} domains (using SITW \textit{dev}). This baseline is referred as \emph{Baseline system S \& T} and we treat it as oracle baseline. 
To compare the performance of \emph{Adaptation system LT} with the system in~\cite{nidadavolu2019cycle}, 
we trained an adaptation system with more \emph{target} domain data (SITW \textit{dev} and VoxCeleb1). 
We refer to this system as \emph{Adaptation system}. Both the adaptation systems in Table \ref{tab:tel_mic_results_baselines} were trained without noise in the \emph{target} domain. 


Results indicate that \emph{Adaptation system LT} suffers in performance compared to \emph{Adaptation system} which was trained on a larger amount of \emph{target} domain data but still improved w.r.t. \emph{Baseline system S}. The performance drop (compared to \emph{Adaptation system}) can either be because of over-fitting or the inability of feature mapping networks to learn good mappings with limited amount of variability in the data.

\begin{table}
	\caption{Results for \emph{Adaptation system LT} trained with noise}
    \label{tab:tel_mic_lt_results}
    \vspace{-2mm}
	\centering
    \resizebox{\columnwidth}{!}{
	\begin{tabular}{@{}lcccc@{}}
    \toprule
    & \multicolumn{2}{c}{\textbf{SITW Core-Core}} & \multicolumn{2}{c}{\textbf{SITW Assist-Multi}} \\
    \cmidrule(lr){2-3}
    \cmidrule(lr){4-5}
    & \textbf{EER} & \textbf{DCF} & \textbf{EER} & \textbf{DCF} \\
    \midrule 
    \textbf{Baseline system S}  & 10.14 & 0.6842 & 12.72 & 0.6941 \\
    \midrule
    \textbf{Adaptation system LT} \\
    $\quad$ without noise & 9.51 & 0.6608 & 11.43 & 0.6683 \\
    \midrule
    \textbf{Adaptation system LT} \\
    $\quad$ with noise & 8.91 & \textbf{0.6495} & \textbf{10.71} & \textbf{0.6608} \\
    \midrule
    \textbf{Adaptation system LT} \\
    $\quad$ with noise on S \& T & 9.98 & 0.6818 & 11.9 & 0.6897 \\
    \midrule
    \textbf{Adaptation system} & \textbf{8.87} & 0.6548 & 10.78 & 0.6643 \\
    \midrule
    \textbf{Baseline system S \& T} & 7.90 & 0.6226 & 10.14 & 0.6418 \\
	\bottomrule
    \end{tabular}}
\end{table}

\subsection{\emph{Adaptation system LT} trained with noise}
\label{ssec:results_of_LT_system_with_noise}
In this section, we experimented with adding noise to the \emph{target} domain data during the training of CycleGAN (procedure mentioned in Section \ref{sec:cg_system}). The intuition behind adding noise is two-fold. One, it is well known that the addition of noise acts as a regularizer during training and prevents over-fitting. Two, the addition of noise to the \emph{target} domain also pulls the distributions apart,
which facilitates better training of GANs. Results are in Table~\ref{tab:tel_mic_lt_results}.

\emph{Adaptation system LT} trained with noise had much better performance compared to \emph{Baseline system S} and slightly better results compared to \emph{Adaptation system}, which was trained with larger \emph{target} domain data. We also experimented with adding noise on both the domains (system referred as \emph{Adaptation system LT} with noise on S \& T) but this system yielded poor results. This justifies our intuition that adding noise on one domain makes the two distributions more different and facilitates better learning for GANs. 
\vspace{-3mm}
\subsection{\emph{Adaptation system} trained with noise}
\label{ssec:results_of_adaptation_system_with_noise}
Here, we investigated the impact of noise addition on the performance of adaptation system when larger amount of \emph{target} domain data was available (SITW \textit{dev} and VoxCeleb1). The motivation was to check whether adding noise to \emph{target} domain would also benefit \textit{non} low-resource adaptation. We trained \emph{Adaptation system} with noise for 50 epochs only as opposed to \emph{Adaptation system} without noise where it was trained for 75 epochs. The results are presented in Table \ref{tab:fmap3}. We observe that both systems yield comparable performance even though system with noise was trained for fewer epochs which indicates addition of noise speeds up the training process. 

\begin{table}[h]
	\caption{Results for \emph{Adaptation system} trained with noise}
    \label{tab:fmap3}

	\centering
    \resizebox{\columnwidth}{!}{
	\begin{tabular}{@{}lcccc@{}}
    \toprule
     & \multicolumn{2}{c}{\textbf{SITW Core-Core}} & \multicolumn{2}{c}{\textbf{SITW Assist-Multi}} \\
    \cmidrule(lr){2-3}
    \cmidrule(lr){4-5}
    & \textbf{EER} & \textbf{DCF} & \textbf{EER} & \textbf{DCF} \\
    \midrule 
    \textbf{Baseline system S}  & 10.14 & 0.6842 & 12.72 & 0.6941 \\
    \midrule
    \textbf{Adaptation System} \\
    $\quad$ without noise (epoch 75)  & 8.87 & \textbf{0.6548} & 10.78 & \textbf{0.6643} \\
    $\quad$ with noise (epoch 50) &  \textbf{8.64 }& 0.6610 & \textbf{10.57} & 0.6682 \\
	\bottomrule
    \end{tabular}}
\end{table}

\vspace{-5mm}
\section{Experimental details for \emph{\MakeLowercase{reverb-clean}} adaptation}
\label{sec:rev_clean_exp_details}
For \textit{reverb-clean} adaptation experiments, we trained the x-vector system on clean data (\emph{source} domain). During evaluation we map features of reverberant speech to clean domain.

\vspace{-3mm}
\subsection{\emph{source} and \emph{target} domain datasets}
\label{ssec:rev_clean_datasets}

A good candidate for the \emph{source} domain dataset should consist of spontaneous speech, have a large number of utterances, and have sufficient speaker variability. It should also be relatively \emph{clean} as it makes the learning to map to a sufficiently different domain more well-defined and effective. To facilitate this, we used VoxCeleb1 and Voxceleb2 \cite{chung2018voxceleb2} for our experiments.  To be able to train on longer audio sequences, we concatenated the files by the same speaker session. This gives us around 2710 hours of spontaneous audio with 7185 speakers. We call this dataset as \emph{voxcelebcat} and it is used for training the x-vector network. Since \emph{voxcelebcat} is collected in wild conditions and may contain unwanted background noise, additional filtering of files is required based on their SNR values to construct the \emph{source} domain data. Inspired from the recent LibriTTS \cite{zen2019libritts} work, we retained only the top 50\% files sorted by their estimated SNR value using WADA-SNR algorithm \cite{kim2008robust}. This gave us our \emph{source} domain dataset which we call \emph{voxcelebcat\_wadasnr}. It consisted of 7104 speakers and duration is around 1665 hours.

We use reverberant versions of SITW \textit{dev} and SITW \textit{eval} as the \emph{target} domain data to train and evaluate the CycleGAN system respectively. The reverberant versions of SITW \textit{dev} and SITW \textit{eval} were created artificially by convolving the original speech signals with real room impulse responses (RIRs) which are publicly available\footnote{\url{http://www.openslr.org/26}}.
The corpus has 315 real RIRs. We randomly selected 285 RIRs from them and used them to corrupt the SITW \textit{dev}. We used the remaining 30 RIRs to make the reverberant copy of SITW \textit{eval}. We refer to both these copies as SITW \textit{dev\_rev} and SITW \textit{dev\_eval} respectively. Similar to \textit{mic-tel} adaptation work we added noise to the \emph{target} domain data (SITW \textit{dev\_rev}) with the same procedure mentioned in section \ref{ssec:tel_mic_noise_addition}. For evaluation, we also used a recent far field VOiCES corpus ~\cite{richey2018voices,nandwana2019voices}. We did not add noise to the evaluation corpora.  All the training and evaluation corpora used in this section was sampled at 16KHz. 
\vspace{-3mm}
\subsection{Baseline and adaptation systems}
\label{ssec:rev_clean_xvec_system}

The x-vector network was trained on \emph{voxcelebcat} dataset mentioned in \ref{ssec:rev_clean_datasets}. 
The features used and other training details were exactly the same as in the \textit{mic-tel} adaptation work. For training the CycleGAN system, we used \emph{voxcelebcat\_wadasnr} and SITW \textit{dev\_rev} with \textit{noise} as the \emph{source} and \emph{target} domain datasets respectively. 
The training details for CycleGAN were the same as in Section \ref{ssec:tel_mic_cg_system}. 
For adaptation system, the features of the evaluation data were mapped to clean domain using the generator $G_{\T\to\S}$. 
These features were forward passed through the x-vector system to get embeddings in the clean domain which were used to score the PLDA model which was trained on clean embeddings from original \emph{source} domain. We also dereverberated the evaluation corpora using the state-of-the-art 
weighted prediction error (WPE)~\cite{nakatani2010speech,yoshioka2012generalization} algorithm and tested the x-vector system on the dereverberted features. We consider it as one of our baselines. 
\vspace{-3mm}

\begin{table}
	\caption{Results for \textit{reverb-clean} adaptation experiments.}
    \label{tab:rev_results}

	\centering
    \resizebox{\columnwidth}{!}{
	\begin{tabular}{@{}lcccc@{}}
    \toprule
    & \multicolumn{2}{c}{\textbf{SITW \textit{eval\_rev} Core}} & \multicolumn{2}{c}{\textbf{VOiCES \textit{eval}}} \\
    \cmidrule(lr){2-3}
    \cmidrule(lr){4-5}
    & \textbf{EER} & \textbf{DCF} & \textbf{EER} & \textbf{DCF} \\
    \midrule 
    \textbf{Baseline system} & 7.24 & 0.5602 & 11.12 & 0.7874 \\
    \textbf{Baseline system with WPE} & 6.52 & 0.5205 & 9.36 & 0.6872 \\
    \midrule
    \textbf{Adaptation system LT} \\
    $\quad$ real RIRs &  6.31 & 0.5018 & 9.77 & 0.7125 \\
    $\quad$ real RIRs and noise & \textbf{6.12} & 0.4974 & 9.08 & \textbf{0.6734} \\
    \midrule
    \textbf{Adaptation system} \\
    $\quad$ real RIRs & 6.26 & 0.5027 &  9.79 & 0.6916 \\
    $\quad$ real RIRs and noise & 6.34 & \textbf{0.4919} & \textbf{8.95} & 0.6780  \\
	\bottomrule
    \end{tabular}}
\end{table}

\section{Results for \emph{\MakeLowercase{reverb-clean}} adaptation}
\label{sec:results_rev_clean}
\vspace{-2mm}
Experimental results are presented in Table \ref{tab:rev_results}. We refer to adaptation system trained with a limited amount of \emph{target} domain data (SITW \textit{dev\_rev} which contains 191 speakers) as \emph{Adaptation system LT}. To compare the performance of this system when a much bigger \emph{target} domain dataset is present during training of CycleGAN we trained a system with reverberant version of \emph{voxcelebcat\_wadasnr} created using the same protocol as SITW \textit{dev\_rev} as the \emph{target} domain data. We name this system \emph{Adaptation system}. All the models are tested on simulated SITW \textit{eval\_rev} created from real RIRs and VOiCES \textit{eval}. To investigate the effect of adding noise, we did our experiments with and without noise. The results demonstrate that adding noise improved the performance, more pronounced improvements were observed on challenging VOiCES \textit{eval} corpus. \emph{Adaptation system LT} with noise and \emph{Adaptation system} yielded comparable performance which is consistent with our observation in \textit{mic-tel} adaptation. Moreover, adaptation systems trained with noise performed slightly better than the state of the art WPE.  \emph{Adaptation system LT} with noise yielded improvements of 18.3\% and 14.5\% relative in terms of EER and DCF compared to \emph{Baseline system} and 3\% and 2\% relative in terms of EER and DCF compared to \emph{Baseline system} with WPE.

\section{Summary and future work}
\label{sec:summary}
We investigated unsupervised domain adaptation using CycleGAN trained with limited amount of \emph{target} domain data. We experimented with two adaptation tasks: microphone to telephone and reverberant to clean speech adaptation to improve the performance of speaker recognition systems trained on \emph{source} domain. We demonstrated that adding noise to \emph{target} domain data while training CycleGAN yielded much better performance compared to a non-adapted baseline system. The low-resource adaptation results are slightly better than the results of an adaptation system trained with a larger amounts of \emph{target} domain data. We also observed noise addition on the \emph{target} domain was beneficial even under non-low-resource setting with an increase in training speed (comparable performance with fewer epochs of training). On the reverberant to clean adaptation task, our low-resource adaptation model yielded 18.4\% and 14.5\% relative improvements compared to a non-adapted baseline system on evaluation corpus of VOiCES and slightly better results compared to the state-of-the-art WPE dereverberation algorithm. In the future, we intend to transfer features from the clean to the reverberant domain using the generator $G_{\S \rightarrow \T}$ and train the x-vector and PLDA backend on the reverberant domain and test it on reverberant speech. We intend to compare this system with an x-vector network trained on augmented data.

\clearpage

\bibliographystyle{IEEEbib}
\bibliography{refs}

\end{document}